\documentclass[12pt]{JHEP3}

\newcommand{\bea}{\begin{eqnarray}}
\newcommand{\beal}[1]{\begin{eqnarray}\label{#1}}
\newcommand{\eea}{\end{eqnarray}} 
\newcommand{\be}{\begin{equation}} 
\newcommand{\bel}[1]{\begin{equation}\label{#1}}
\newcommand{\ee}{\end{equation}}

\newcommand{\bit}{\begin{itemize}}
\newcommand{\eit}{\end{itemize}}
\newcommand{\ben}{\begin{enumerate}}
\newcommand{\een}{\end{enumerate}}

\def\half{\frac{1}{2}}

\def\alp{\leavevmode\ifmmode {\alpha^\prime} \else ${\alpha^\prime}$ \fi}
\def\mps{M_P^2}
\def\fnl{f_{NL}}
\def\nnl{n_{NL}}
\def\froot{\sqrt{1+\frac{108}{35}\fnl}}

\title{A Consistency Relation for Power Law Inflation in DBI Models}

\preprint{}

\author{Micha\l\ Spali\'nski\footnote{Email: mspal@fuw.edu.pl}\\
So\l tan Institute for Nuclear Studies\\
ul. Ho\.za 69, 
00-681 Warszawa, Polska.
}

\abstract{Brane inflation in string theory leads to a new realization of
power law inflation which can give rise to significant
non-gaussianity. This can happen for any throat geometry if the scalar
potential is appropriate.  This note presents a consistency relation
connecting the running of the nonlinearity parameter characterizing the
non-gaussianity and the scalar and tensor indices. The relationship is
valid assuming that the throat geometry and scalar potential support power
law inflation, regardless of the level of non-gaussianity.}


\begin{document}

\section{Introduction}

Given the importance of inflation in our current view of cosmology it is
natural and important to try to understand the details of it in the
framework of string theory. 
From the string theory perspective general relativity is a low energy
effective field theory, which 
receives corrections both at the classical and quantum level. These
corrections may be crucial in the very early stages of evolution of our
Universe. 
String theory should also
determine the degrees of freedom relevant at the time inflation is expected
to occur; specifically, it should provide an inflaton. 

One possible scenario is brane inflation
\cite{Dvali:1998pa}--\cite{Thomas:2007sj}, which interprets inflation as
the motion of a $D3$-brane down a throat in a warped Calabi-Yau
compactification \cite{Giddings:2001yu}. Brane inflation has a rather
distinct character because the inflaton is identified with a brane
position, and the relevant scalar field kinetic energy functional is
non-canonical. Its form is determined by T-duality to be a
Dirac-Born-Infeld action \cite{Leigh:1989jq}.  When restricted to spatially
homogeneous configurations the action of the inflaton reduces to a DBI
scalar field theory studied in a number of papers over the past few years
\cite{Silverstein:2003hf}--\cite{Spalinski:2007dv}.  The difference between
the DBI action and a canonical scalar field action may be interpreted as a
classical correction coming from string theory.
Brane inflation also introduces a new interpretation for the end of
inflation and the thermalization of standard model degrees of
freedom: these phenomena come about in consequence of brane annihilation.  

It is clearly very important to try to determine observational
possibilities which could distinguish this scenario from other options.
From the point of view of comparing inflationary models to observation
there are a number of properties which significantly restrict the spectrum
of possibilities. In this context the most important quantities are
inflationary observables like the scalar and tensor spectral indices and
the primordial non-gaussianity, as well as the running of these quantities.

One of the interesting features of DBI inflation is the natural appearance
of significant levels of primordial non-gaussianity in the spectrum of
curvature perturbations. This is a direct consequence of the nonlinear
corrections to the scalar kinetic terms.  Furthermore, as Chen has
emphasized \cite{Chen:2005fe}, deviations from scale invariance responsible
for the running of the scalar and tensor spectral indices also induce
running of the non-gaussianity. The current observational limits 
allow quite significant levels of non-gaussianity and it will be very
interesting to see whether it will turn out to be non-vanishing. At the
moment one has to regard it as an important dimension of the inflationary
parameter space. In terms of the non-gaussianity parameter $\fnl$ the
current limits \cite{Creminelli:2005hu} give $|\fnl|<300$. 
Single field inflation models give $|\fnl|\approx 1$ at most
\cite{Creminelli:2003iq}, so if coming 
experiments observe $|\fnl|$ of the order of a few or more it will indicate
that models of inflation based on a single canonical field need to be
extended either by allowing nontrivial dynamics (as in DBI models) or by
having multiple scalars evolving during the inflationary stage. 

In an interesting recent paper on brane inflation \cite{Lidsey:2006ia}
Lidsey and Seery have derived a rather general relation involving the
nonlinearity parameter often used to describe primordial non-gaussianity in
certain simple kinematical configurations.  This note applies the same
approach to power law inflation in DBI scalar field theories. The result is
a relation between the inflationary observables involving the running of
the nonlinearity parameter. The observational prospects for actually
measuring this quantity are remote, but perhaps not hopeless.

\section{Inflationary Observables}
\label{obsect}

Inflationary observables related
to the primordial perturbation spectra have been calculated  
(to leading order in the Hubble slow roll parameters) by Garriga and
Mukhanov \cite{Garriga:1999vw} for a wide
class of scalar field theories, which can be described by the action 
\bel{genact}
S = \int d^4 x \sqrt{-g} (R + P(X,\phi)) \ ,
\ee
where $X\equiv -\half (\partial\phi)^2$. Their results 
can be written in terms of Hubble slow roll parameters 
\bea
\epsilon_H &=& -\frac{1}{H} \frac{d}{dt} \ln H \label{hepsilon}\\
\eta_H &=& -\frac{1}{H} \frac{d}{dt} \ln \epsilon_H \label{heta}\\
\sigma_H &=& -\frac{1}{H} \frac{d}{dt} \ln c_s \label{hsigma}\ ,
\eea
where $c_s^2=p_{,X}/\rho_{,X}$ is the speed of sound:
\be
c_s^2 = \frac{P_{,X}} {P_{,X} + 2 X P_{,XX}} \ .
\ee
The spectral indices are then given by 
\bea
n_S-1&=&-2\epsilon_H + \eta_H + \sigma_H \label{ns}\\
n_T &=& -2 \epsilon_H \ \label{nt} \ .
\eea
These expressions are valid in the leading order in Hubble slow roll
parameters (\ref{hepsilon})-(\ref{heta}), which are assumed to be small
during the observable phase of inflation. 

For the sequel one also needs to recall the notion of the ``nonlinearity''
parameter $\fnl$, which is an often used measure of
non-gaussianity\footnote{See for example \cite{Seery:2005gb,Chen:2006nt}
  for the precise definition.}.    
A simple and explicit formula, valid in a wide range of scalar field
theories defined by (\ref{genact}), 
has recently been obtained in \cite{Chen:2006nt}: 
\bel{fnl}
\fnl = \frac{35}{108} (\frac{1}{c_s^2} - 1) - 
\frac{5}{81} (\frac{1}{c_s^2} - 1 - 2\Lambda)
\ee
where 
\bel{biglam}
\Lambda \equiv \frac{X^2 P_{,XX} + \frac{2}{3} X^3 P_{,XXX}}{X P,X + 2X^2
  P,XX} \ .
\ee
As emphasized by Chen \cite{Chen:2005fe}, deviations from scale invariance
should manifest themselves also in the 
running of the non-gaussianity. A measure of it is the index
\bel{nnl}
\nnl \equiv \frac{d\ln\fnl}{d\ln k}
\ee
defined in \cite{Chen:2005fe}.

\section{DBI scalar field theories}
\label{dbisect}

The inflaton in brane inflation scenarios is an open string mode, 
which implies that its dynamics are described by 
the Dirac-Born-Infeld action. For spatially homogeneous inflaton
configurations the action takes the form
\cite{Silverstein:2003hf,Shandera:2006ax}  
\bel{dbi}
S = - \int d^4x\ a(t)^3\ \{f(\phi)^{-1}(\sqrt{1-f(\phi) \dot{\phi}^2}-1) +  
V(\phi)\}  \ .
\ee
The function $f$ appearing here can be expressed in terms of the warp
factor in the metric and the $D$3-brane tension. The function $f$
appearing here is positive by construction\footnote{Similar actions with
  negative  $f$ have also been discussed in the
  literature\cite{Mukhanov:2005bu}--\cite{Babichev:2006vx}.}.  

It is convenient to use the Hamilton-Jacobi
formalism \cite{Markov:1988yx}--\cite{Kinney:1997ne}, 
which makes use of the Hubble parameter expressed as function
of the scalar field\footnote{In the context of DBI scalar field theories
  the Hamilton-Jacobi formalism was introduced in
  \cite{Silverstein:2003hf} and was recently discussed in
  \cite{Spalinski:2007kt}.}.  
The basic point is to eliminate the field derivative using the relation 
\bel{phidot}
\dot{\phi} = - \frac{2\mps}{\gamma} H' \ ,
\ee
where $\gamma$ is given as a function of $\phi$ by
\bel{gamma}
\gamma(\phi)=\sqrt{1+ 4 M_P^4 f(\phi) H'(\phi)^2} \ .
\ee
Using this one can calculate the Hubble slow roll parameters
\bea
\epsilon_H &=& \frac{2\mps}{\gamma} (\frac{H'}{H})^2 \label{dbieps} \\
\sigma_H &=& -\frac{2\mps}{\gamma} \frac{H'}{H}  \frac{\gamma'}{\gamma} 
\label{dbisig} \\
\eta_H &=& \frac{4\mps}{\gamma} \frac{H''}{H} -2\epsilon_H + \sigma_H
\label{dbieta} \ .
\eea
The formula for the nonlinearity
parameter (\ref{fnl}) in the present case simplifies, since one has
\cite{Chen:2006nt} $c_s=\gamma^{-1}$ and $\Lambda=0$. This leads to the
simple result \cite{Alishahiha:2004eh}, \cite{Chen:2006nt}: 
\bel{dbifnl}
\fnl = \frac{35}{108} (\gamma^2 - 1) \ .
\ee
This shows that non-gaussianity in DBI models becomes large in the
``ultra-relativistic'' regime $\gamma\gg 1$ \cite{Alishahiha:2004eh}.  

Lidsey and Seery \cite{Lidsey:2006ia} have 
noted that using (\ref{nt}) and (\ref{dbifnl}) one can turn the expression
for the tensor to scalar ratio 
\bel{dbir}
r = \frac{16\epsilon_H}{\gamma}
\ee
into a consistency relation involving only observable
parameters:
\bel{fsrel}
8 n_T = -r \froot \ ,
\ee
which is valid for any DBI scalar field theory, and generalizes the
usual consistency relation appearing in \cite{Lidsey:1995np}. It is a very
interesting, testable, prediction of the brane inflation scenario. 

The authors of \cite{Lidsey:2006ia} also considered a special case of
inflation near 
the bottom of a warped throat \cite{Kecskemeti:2006cg} to derive further
relations between 
observable parameters in that situation. 
In a similar spirit, the following section turns to
power 
law inflation in DBI scalar field theories, where one can obtain another 
consistency relation of this type, involving the running non-gaussianity
parameter (\ref{nnl}), which can easily be calculated in this class 
of models: 
\bel{dbinnl}
\nnl = -4\mps (1+\frac{35}{108}\frac{1}{\fnl}) \frac{\gamma'}{\gamma}
\frac{H'}{H} 
\ee
This is valid to leading order in the Hubble slow roll parameters.
As explained in the following section, if one assumes power
law inflation then using (\ref{dbinnl}) it is possible
to rewrite (\ref{ns}) as another consistency relation.

\section {Power Law Inflation}
\label{powersect}

It was found by Silverstein and Tong \cite{Silverstein:2003hf} that for the
case of an AdS throat (where $f(\phi) = \lambda/\phi^4$) a quadratic
potential with a suitably high 
inflaton mass leads to power law inflation\footnote{Other realizations of
  power law inflation in string theory are described in \cite{Becker:2005sg}
  and \cite{Dimopoulos:2005ac}.} in the ``ultra-relativistic''
regime $\gamma\gg 1$. It was subsequently pointed out that power law
inflationary solutions exist in DBI scalar field theories even when
$\gamma$ is not large \cite{Spalinski:2007dv}. Furthermore, for any throat
geometry there is a 
potential which leads to power law inflation for some range of
parameters. This generalizes the well known fact that in the case of
canonical kinetic terms exponential potentials lead to power law
inflation \cite{Lucchin:1984yf}. 

Power law inflation occurs when 
the parameter $w$ in the baryotropic equation of
state $p=w\rho$ is constant and $w<-1/3$. As shown in
\cite{Spalinski:2007dv}, power law inflationary solutions will
exist if the potential is of the form 
\bel{hjdbi}
V(\phi) = 3\mps H(\phi)^2 - \frac{\gamma(\phi)-1}{f(\phi)} \ ,
\ee
where $H(\phi)$ satisfies the differential equation 
\bel{bardbi}
4\mps H'^2 = 3 (w+1) H^2 \sqrt{1 + 4 M_P^4 f H'^2} 
\ee
with $w<-1/3$. 

The essential property of power law inflation is that the parameter
$\epsilon_H$ is constant.
Indeed, from (\ref{dbieps}) and (\ref{bardbi}) one concludes that  
\be
\epsilon_H = \frac{3}{2} (w+1). 
\ee
One immediate consequence is that the tensor spectral index does not run,
since by virtue of (\ref{nt}) it is constant. Furthermore, a measurement
of the tensor spectral index would  
determine the parameter $w$, which in the model of
\cite{Silverstein:2003hf} is  
related to the inflaton mass \cite{Spalinski:2007dv}. 

Since $\epsilon_H$ is constant it also follows that $\eta_H$ (defined
in (\ref{heta})) vanishes\footnote{Some authors
  (e.g. \cite{Shandera:2006ax}) define a 
  different  ``$\eta$'' parameter in this context, related to $\eta_H$ by
  $2\eta_D=\eta_H+2\epsilon_H-\sigma_H$. In that language power law inflation
  implies the relation $\eta_D=\epsilon_H-\sigma_H/2$.}.
This makes it possible to 
derive another consistency relation involving the spectral indices, the
non-gaussianity, and running of the nonlinearity parameter (\ref{nnl}). 
Indeed, from (\ref{dbisig}) and (\ref{dbinnl}) it follows that  
\be
\sigma_H = \half \nnl (1+\frac{35}{108}\frac{1}{\fnl})^{-1} \ .
\ee
Using this and $\eta_H=0$ in (\ref{ns}), (\ref{nt}) one finds
\bel{powercons}
n_S - n_T = 1 + \half \nnl (1+\frac{35}{108}\frac{1}{\fnl})^{-1} \ ,
\ee
which is a relation between observable parameters. It is valid for any DBI
scalar field theory solution describing power law inflation. In particular,
it does not assume simplifications which occur in the
``ultra-relativistic'' limit, so one can also consider the case of $\fnl$
small or zero in this expression. This implies, in particular, that power
law inflation with canonical kinetic terms has $n_S - n_T = 1$.  

In the ``ultra-relativistic'' limit, when the non-gaussianity is large, this
relation can be further simplified to
\be
n_S - n_T = 1 + \half \nnl \ .
\ee
While the prospect of measuring $\nnl$ seems distant today, these 
relations may be tested at some point in the future.

\section{Conclusions}

Single field inflation with canonical kinetic energy terms 
leads to negligible non-gaussianity
\cite{Maldacena:2002vr}, \cite{Acquaviva:2002ud}. While it is too 
early to tell whether observation will require more general models of
inflation, a lot of attention has been devoted to models where large
non-gaussianity may naturally occur. One possibility is models with
multiple scalars \cite{Seery:2005gb}. Another option is DBI inflation,
which can generate significant non-gaussianity during 
a power law inflationary stage. 

In field theoretical models power law inflation is realized by an
exponential scalar potential, so there is no natural mechanism for
inflation to end. One needs to supplement the 
exponential potential by some external agent which terminates inflation. In
the 
context of brane inflation this role is played by a tachyon 
which appears when the 
mobile $D3$-brane gets within a warped string length of the
anti-brane at the bottom of the throat. The process of brane annihilation
ends inflation and the energy released is 
(hopefully \cite{Kofman:2005yz}, \cite{Frey:2005jk}) 
transferred to  standard-model degrees of freedom\footnote{This process
  could in fact be quite complex, as recently emphasised in
  \cite{Allahverdi:2006iq}, \cite{Allahverdi:2007wh}.}   
localized in another throat in the  
compactification manifold. One may thus argue that power law inflation
finds a very natural 
place in the brane inflation scheme. 

The consistency relation (\ref{powercons}) is a
consequence of assuming power law inflation, but it is valid in 
DBI scalar field theories without necessarily assuming the
``ultra-relativistic'' limit $\gamma\gg 1$. It is also worth stressing that
it is not 
restricted to the specific realization of power law inflation discussed in
\cite{Silverstein:2003hf}, i.e. a quadratic potential and an anti-de-Sitter
throat\footnote{As discussed in reference \cite{Bean:2007hc}, this case is
  probably already ruled out by observation. }.   
This is rather important, in that there are many contributions to the
scalar potential, which are at the moment hard to control. There are also
various possibilities for warped throats in type IIB
compactifications, and different opinions as to which section of
the throat is relevant for inflation \cite{Kecskemeti:2006cg}, as well as
to the direction 
of the $D$-brane motion \cite{Chen:2004hu}, \cite{Chen:2005ad}. 
The consistency relation derived here does not assume a specific choice in
these matters; it should be valid whenever the resulting inflationary stage
has power law character.


\end{document}